\documentclass[trackchanges,twocolumn]{aastex701}
\usepackage{amsmath}
\usepackage{soul}
\begin{document}

\title{Implications for Primordial Black Hole Dark Matter from a Single Subsolar Mass
Gravitational-wave Detection in LVK O1--O4}

\author[orcid=0009-0009-6060-9540,gname='Alberto' , sname='Magaraggia']{Alberto Magaraggia}
\affiliation{Department of Physics, University of Miami, Coral Gables, FL 33124, USA}
\email{axm8568@miami.edu}
\author[orcid=0000-0002-1697-186X,sname='Nico, gname=Cappelluti']{Nico Cappelluti}
%\altaffiliation{University of Miami}
\affiliation{Department of Physics, University of Miami, Coral Gables, FL 33124, USA}
\email[show]{ncappelluti@miami.edu}

%% Use the \collaboration command to identify collaborations. This command
%% takes an optional argument that is either a number or the word "all"
%% which tells the compiler how many of the authors above the command to
%% show. For example "\collaboration[all]{(DELVE Collaboration)}" wil include
%% all the authors above this command.
%%
%% Mark off the abstract in the ``abstract'' environment. 
\begin{abstract}
The detection of sub-solar mass black holes is a milestone of modern astrophysics as it would open a window either onto new stellar physics or could potentially unveil the nature of Dark Matter as Primordial Black Holes (PBHs). On November 12, 2025, the LIGO-Virgo-KAGRA (LVK) collaboration reported the compact binary merger candidate S251112cm, a system with no obvious EM counterpart, consistent with binary black hole merger with a chirp mass in the range $0.1-0.87 M_\odot$. The probability that at least one component has mass $<$1 $M_{\odot}$ is $>99\%$. Inspired by this trigger, we tested if a population of PBHs formed at Quantum Chromodynamics epoch with a broad mass function could account for a signal of this type. Our results, corresponding to a predicted event rate of $0.8 \,\text{yr}^{-1}$ as seen by LVK O3b, suggest that the observed merger rate of $0.23^{+0.86}_{-0.218}\,\text{yr}^{-1}\;(95\%\;\text{C.L.})$ if the trigger is confirmed as an astrophysical event would be compatible with such a model. Our predicted detection rate is also in agreement with current LVK expectations for stellar-mass binaries, remaining consistent with a scenario in which a non-negligible fraction of the $3-200 \;M_\odot$ mergers observed by LVK originate from Primordial Black Holes. If confirmed, this detection would place a lower limit to the PBH abundance $f_{PBH}>0.04$ for our adopted model. 
\end{abstract}

%% Keywords should appear after the \end{abstract} command. 
%% The AAS Journals now uses Unified Astronomy Thesaurus (UAT) concepts:
%% https://astrothesaurus.org
%% You will be asked to selected these concepts during the submission process
%% but this old "keyword" functionality is maintained in case authors want
%% to include these concepts in their preprints.
%%
%% You can use the \uat command to link your UAT concepts back its source.

%% From the front matter, we move on to the body of the paper.
%% Sections are demarcated by \section and \subsection, respectively.
%% Observe the use of the LaTeX \label
%% command after the \subsection to give a symbolic KEY to the
%% subsection for cross-referencing in a \ref command.
%% You can use LaTeX's \ref and \label commands to keep track of
%% cross-references to sections, equations, tables, and figures.
%% That way, if you change the order of any elements, LaTeX will
%% automatically renumber them.

\section{Introduction} 
The observational landscape of compact binary mergers, dramatically reshaped by the LIGO-Virgo-KAGRA (LVK) collaboration, continues to present challenges to traditional astrophysics. A critical region of interest is the sub-solar mass range ($M \lesssim 1 M_{\odot}$), a region where compact objects are not expected to form through conventional stellar collapse mechanisms. The discovery of binary black hole (BBH) systems with components residing in this mass regime would strongly support a non-stellar origin, most prominently the formation of Primordial Black Holes \citep[PBHs,][]{hawking71,carr74,1975Natur.253..251C} in the early Universe. If PBHs exist, they could account for part or all the Dark Matter (DM) content of the Universe.

On November 12, 2025, the LVK collaboration reported the compact binary merger candidate S251112cm \citep{S251112cm}. This event is statistically compelling due to its relatively low False Alarm Rate (FAR), initially estimated at $\sim 1$ per 6 years, and subsequently refined to $\sim 1$ per 4 years \citep{2025GCN.42690....1L} by the MBTA SSM pipeline. The event's source luminosity distance is estimated to be $93 \pm 27\ \mathrm{Mpc}$, and the detection by the LIGO Hanford (H1), LIGO Livingston (L1), and Virgo (V1) interferometers resulted in a large $90\%$ credible sky-localization region of $1681\ \mathrm{deg}^2$. The analysis indicated that the source chirp mass ($\mathcal{M}_c$) falls predominantly in the range [0.1, 0.87] $M_{\odot}$, and the probability that the lighter component is a sub-solar mass object (HasSSM) is ${>99\%}$. This low-mass characteristic places S251112cm firmly within the predicted mass signature of PBHs. Thus, this trigger serves as a useful motivation to investigate the implications of a confident detection of a sub-solar gravitational-wave (GW) event on PBH mass functions.

Several works suggest that PBHs formed during the Quantum Chromodynamics (QCD) epoch \citep[see e.g.,][]{2017JPhCS.840a2032G,carr2020,Bodeker2021-hz}, resulting in an extended PBH mass function (expressed as $df_{PBH}/d\ln M$) that peaks in the sub-solar to solar mass range and extending from the asteroid to the Supermassive BH range. 

The  focus of this paper is to demonstrate the quantitative consistency of one sub-solar GW detection by LVK with the GW signature predicted by QCD-epoch PBH mass function \citep{2017JPhCS.840a2032G,Hasinger2020,Bodeker2021-hz}, deferring a detailed study of alternative astrophysical models.

The results presented here are based on the public LVK alert S251112cm. Although its reported False Alarm Rate is approximately one per four years, the corresponding offline analysis and parameter-estimation results have not yet been released publicly, and the alert may still be revised or retracted once the final analysis becomes available.
We also emphasize that the reported chirp-mass interval reflects the design of the dedicated sub-solar-mass search that generated the alert. The astrophysical significance therefore lies not in the numerical value of the mass estimate per se, but in the implication that, if the event is confirmed, nature may have produced a compact binary in a mass regime not populated by standard stellar-remnant black holes.

Throughout this paper we assume Planck 2018 Cosmology \citep{Planck_Collaboration2020-ij} with $H_0 = 67.6 \mathrm{\ km\,s^{-1}\,Mpc^{-1}}$, a matter density parameter $\Omega_m$= 0.311 and $\Omega_{\Lambda}$=0.689.

%% This command is needed to show the entire author+affiliation list when
%% the collaboration and author truncation commands are used.  It has to
%% go at the end of the manuscript.
%\allauthors

%% Include this line if you are using the \added, \replaced, \deleted
%% commands to see a summary list of all changes at the end of the article.
%\listofchanges
\section{Extended PBH mass function}
Primordial Black Holes probe high-energy physics and the early thermal history of the Universe, as their mass spectrum records variations in the equation of state during the radiation dominated phase. Phase transitions such as QCD confinement and particle--annihilation epochs briefly soften the equation of state, enhancing PBH production at the corresponding horizon masses and generating a multimodal, physically motivated distribution \citep{Hasinger2020}. Early lepton-flavour asymmetries can further modify the relativistic degrees of freedom near the QCD epoch, leaving an imprint in the PBH mass function \citep{Bodeker2021-hz}.

\begin{figure*}
    \centering
    \includegraphics[width=\linewidth]{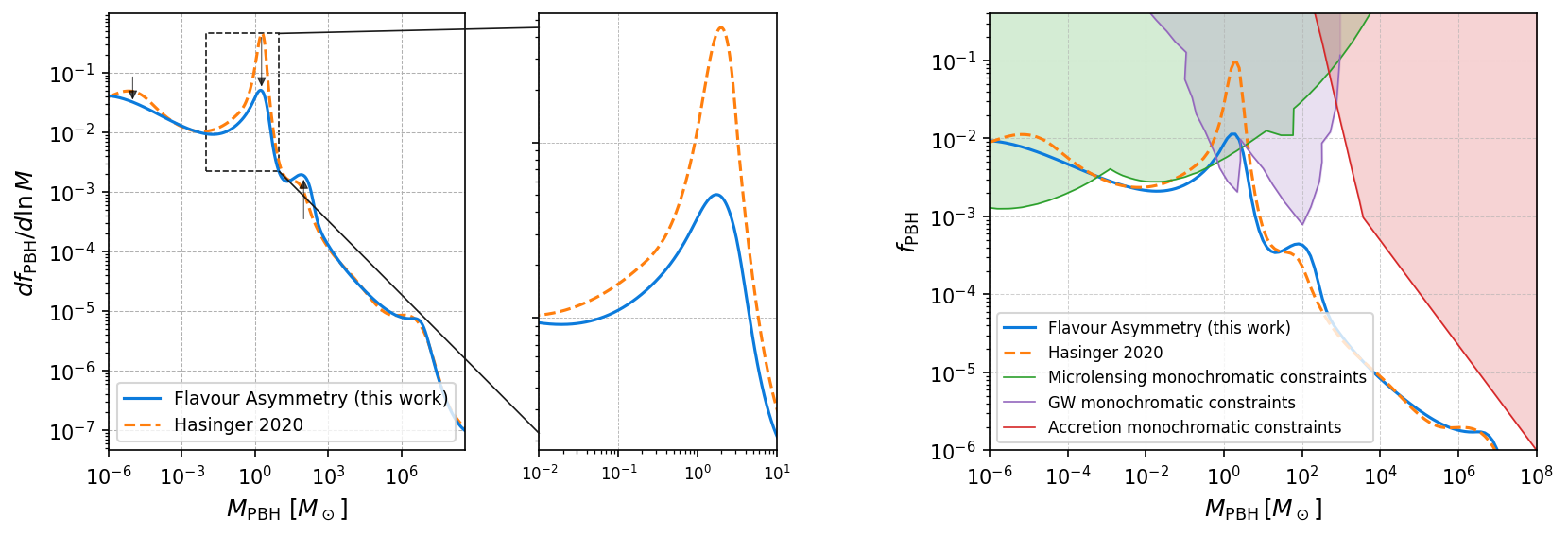}
    \caption{Left panel: Mass function expressed as \(d f_{\rm PBH}/d\ln M\). The orange dashed line shows the distribution from \citep{Hasinger2020}, while the blue solid line represents the same model modified by \cite{Bodeker2021-hz} to account for lepton--flavour asymmetries. The three black arrows highlight the main changes introduced by these asymmetries. Right panel: rough comparison between the abundances \(f_{\rm PBH}\) associated with the two mass functions and the current monochromatic constraints (red: CMB-accretion limits, green: microlensing bounds, purple: gravitational-wave limits).}
    \label{fig:massfunction}
\end{figure*}

To model these effects, we adopt the extended PBH mass spectrum of \citet{Hasinger2020}, which produces four PBH populations across 
$[10^{-6}M_\odot\;,\;4\times10^{8}M_\odot]$. Planetary-mass PBHs ($\sim10^{-5}M_\odot$) form at the electroweak transition; PBHs near the Chandrasekhar scale ($\sim1.5M_\odot$) arise when baryons appear; PBHs of $\sim50M_\odot$ emerge during pion formation; and supermassive PBHs ($\sim10^{7}M_\odot$) originate around $e^{+}e^{-}$ annihilation, producing the peaks shown by the dashed orange curve in Fig.~\ref{fig:massfunction}.

Lepton-flavour asymmetries,
$\ell_\alpha = \frac{n_\alpha - n_{\bar{\alpha}} + n_{\nu_\alpha} - n_{\bar{\nu}_\alpha}}{s}$,
represent the excess of leptons and neutrinos over their antiparticles and induce non-zero chemical potentials. These shift the effective equation-of-state parameter $w(T,\ell_\alpha)$ during epochs such as the QCD and pion/muon transitions, modifying collapse thresholds and imprinting new features in the PBH spectrum. Current bounds allow $|\ell_e+\ell_\mu+\ell_\tau| < 1.2\times 10^{-2}$ \citep{Bodeker2021-hz}.

We incorporate the asymmetries $\ell_e=-8\times10^{-2}$ and $\ell_\mu=\ell_\tau=4\times10^{-2}$ \citep{Bodeker2021-hz}, which reduce the $\sim10^{-5}M_\odot$ population, strongly suppress the $\sim1.5M_\odot$ peak, and generate a minor enhancement near $100\,M_\odot$, as indicated by the arrows in the first panel of Fig.~\ref{fig:massfunction}. The resulting PBH spectrum used in this work is shown by the solid blue curve.

Throughout the analysis we assume, in the mass range considered here ($[10^{-6},4\times10^8]M_\odot$), a total PBH dark-matter fraction 
$f_{\rm PBH}=\int \frac{{\rm d}f_{\rm PBH}}{{\rm d}M}\,{\rm d}M = 0.339$,
yielding a scenario in which PBHs form a significant, physically motivated component of the dark matter. It is worth noticing that while in the mass range considered here f$_{PBH}<1$, the model predicts that the bulk of the dark matter lies in the sub-planetary/asteroid mass range ($M<10^{-6}\,M_\odot$), and therefore, this model can easily account for the totality of Dark Matter across more than 20 orders of magnitude in BH mass.

\subsection{Constraints on Primordial Black Holes}
Primordial Black Holes are constrained by a wide set of astrophysical and cosmological observations that probe their impact across many orders of magnitude in mass \citep{Carr2025}. The strongest limits arise from the evaporation of very low--mass PBHs, microlensing surveys, dynamical heating in stellar systems, CMB energy injection from accretion, and gravitational-wave observations. Together, these constraints delimit the fraction of dark matter in PBHs, $f_{\rm PBH}(M)$, over essentially the entire mass spectrum. Most published bounds, however, are derived under the simplifying assumption of a \emph{monochromatic} mass distribution, in which all PBHs share the same mass $M$. These monochromatic constraints provide a useful benchmark but do not capture the phenomenology of extended mass functions.

As emphasized by \cite{Bellomo2018-fz}, one must account for the mass dependence of each observable by performing a convolution over the extended mass function. An effective approach is to map the extended distribution to an ``equivalent'' monochromatic mass that reproduces the same physical effect for a given observable. Only after this step can monochromatic limits be consistently reinterpreted. Without this additional calculation, comparing extended mass functions to published monochromatic bounds is generally unreliable.

In the present work we do not attempt a full conversion but include, for completeness, a rough visual comparison with current monochromatic constraints from CMB accretion, microlensing, and gravitational waves (Fig. \ref{fig:massfunction}, right panel). We approximate the abundance in a finite mass bin by
\(f_{\rm PBH}(\bar M) \simeq \left.\frac{{\rm d}f_{\rm PBH}}{{\rm d}\ln M}\right|_{\bar M}\,\Delta\ln M\),
where \(\bar M\) is the representative mass of the bin and
\(\Delta\ln M = 0.1\,\ln 10\) is set by the grid spacing of the mass function.
This figure is intended only as an orienting reference: quantitative statements for extended mass functions require the full procedure outlined in \cite{Bellomo2018-fz}.

\section{Merger rate estimate}
\label{sec:merger_rate}
\subsection{Formation of PBH binaries}
Primordial Black Holes, formed at the beginning of the Universe, may have become bound into pairs through two separate mechanisms. They could have formed PBH binaries in the early Universe by decoupling from the Hubble flow \citep{Raidal_2019}, or in the late Universe by gravitational capture \citep{https://doi.org/10.48550/arxiv.1610.08479}. Once bound, two objects in a binary system will shrink their orbit emitting gravitational waves, until they eventually merge, releasing gravitational radiation which can be detected through the current interferometers.

Primordial Black Holes are believed to be formed without initial clustering \citep{Dizgah_2019}, so immediately after their production they are sparsely distributed, with a mean comoving separation much larger than the Hubble horizon. During the radiation--dominated epoch the physical separation between PBHs grows as $a(t)\propto t^{1/2}$, while the Hubble horizon $r_H\sim H^{-1}(t)$ increases linearly with time. 
Consequently, the ratio between the PBH mean distance and the horizon scale decreases as the Universe expands. As this ratio falls below unity, typical Hubble patches begin to enclose more than one PBH, allowing close pairs to decouple from the cosmic expansion and form bound systems. 
Once a pair becomes gravitationally bound, nearby PBHs generate tidal torques that give the binary a nonzero angular momentum, preventing a head--on collision and driving the formation of an eccentric PBH binary \citep{Sasaki_2018}.

Consider now a scenario in the late Universe, after structure formation, in which a PBH moving through space experiences a close encounter with another PBH, for example in a region where PBHs are locally concentrated inside a larger dark matter halo. Let the two objects approach each other with relative velocity at infinity $v$. A bound system can form if the gravitational--wave energy emitted during the encounter is sufficient to compensate the initial kinetic energy, namely if $E_\infty \le |\Delta E_{\rm GW}|$ \citep{https://doi.org/10.48550/arxiv.1610.08479}. This condition imposes an upper limit on the allowed impact parameter, defining a maximal value $b_{\max}$ below which the encounter leads to binary capture.

Both formation mechanisms can in principle coexist. However, in this paper we will focus on the late PBH binaries formation channel, since it is characterized better and it is expected to dominate. The fraction of binaries
decoupled in the early Universe is expected to be of the order of $\mathcal{O}\!\left(0.01\, f_{\rm PBH}^{16/37}\right)$ \citep{PhysRevD.103.023026_wong}: even for $f_{PBH} = 1$ the
formation of binaries in the early Universe is expected to be a subdominant process.

\subsection{Late PBH binaries merger rate density}

The cosmic--mean dark--matter density is $\rho_{\rm DM} = \Omega_{\rm DM}\,\rho_c$. Galaxy haloes, where we assume most of PBHs live in, have a local overdensity $\delta_{\rm loc}$ compared to the mean value, such that $\rho_{\rm DM}^{\rm loc} = \delta_{\rm loc}\,\rho_{\rm DM}\,$. The exact value of $\rho_{\rm DM}^{\rm loc}$ is uncertain, but for Milky Way-like galaxies, it has been estimated to be $\rho_{\rm DM}^{\rm loc} \simeq 0.3-0.5 \;\mathrm{GeV cm^{-3}}$ \citep{deSalas_2021} from most global analyses. We will assume $\rho_{\rm DM}^{\rm loc}= 0.4 \;\mathrm{GeV cm^{-3}}$ resulting in $\delta_{\rm loc} \sim 3\cdot10^5$.
Moreover, we will consider the PBH inside the halos to move at the most likely DM velocity in Milky Way-like galaxies: the velocities usually follow a Maxwell--Boltzmann distribution, peaked at around $v_{vir}=250 \;\mathrm{km \;s^{-1}}$ \citep{folsom2025darkmattervelocitydistributions}. In this paper, we will use $v=\sqrt{2}\;v_{vir}$ \citep{PhysRevD.94.084013} as the typical relative velocity of two PBH.

We can express the comoving number density of PBH per logarithmic mass interval as \citep{Bellomo2018-fz}
\begin{equation}
    \frac{{\rm d}n}{{\rm d}\ln M}
    = \ln10 \,\frac{\rho_{\rm DM}^{\rm loc}}{M} \,
      \frac{{\rm d}f_{\rm PBH}}{{\rm d}\ln M} \,,
\end{equation}
so that in a finite logarithmic bin of width $\Delta\ln M$ one has
\begin{align}
\begin{split}
    & n(M_1) \simeq 
    \left.\frac{{\rm d}n}{{\rm d}\ln M}\right|_{M_1}\,\Delta\ln M
    = \\  
    &= \ln10 \,\frac{\rho_{\rm DM}^{\rm loc}}{M_1}\,
    \left.\frac{{\rm d}f_{\rm PBH}}{{\rm d}\ln M}\right|_{M_1}
    \,\Delta\ln M \, = \ln10 \,\frac{\rho_{\rm DM}^{\rm loc}}{M_1} f_{PBH,1}.
\end{split}
\end{align}
For a pair with component masses $M_1$ and $M_2$, we denote the total mass as $M_{tot} = M_1 + M_2$.
In the gravitational--wave capture channel, the capture cross section of two PBHs with relative velocity $v$ is then \citep{https://doi.org/10.48550/arxiv.1610.08479}:
\begin{align}
\begin{split}
    &\bar{\sigma}(M_1,M_2;v)
    = \pi \;b_{max}^2 =\\
    &=2\pi \,
      \left(\frac{85\pi}{6\sqrt{2}}\right)^{2/7}
      G^2\,M_{tot}^{10/7}(M_1 M_2)^{2/7}
      c^{-10/7} v^{-18/7} \,,
\end{split}
\end{align}
and the intrinsic merger rate density for a given pair $(M_1,M_2)$ is:
\begin{align}
\begin{split}
    &\tau_L(M_1,M_2)
    = n(M_1)\,n(M_2)\,\bar{\sigma}(M_1,M_2;v)\,v \,= \\
    &= 2\pi \;(\ln10)^2\left(\frac{85\pi}{6\sqrt{2}}\right)^{2/7}
\frac{G^{2}\,\rho_{\rm DM}^{2}\,\delta_{\rm loc}^{2}\,f_{\rm PBH,1}f_{\rm PBH,2}}{c^{10/7}\,v^{11/7}}\,\times \\
&\times\frac{M_{tot}^{10/7}}{(M_1 M_2)^{5/7}}\,
.
\label{eq:tau_L}
\end{split}
\end{align}

The late--Universe PBH merger rate adopted in this work should be regarded as intrinsically uncertain and not a robust prediction, although it remains one of the most commonly used phenomenological prescription in the literature for PBH binaries forming after structure formation. Recent N-body studies highlight this uncertainty: simulations of small, isolated PBH clusters without primordial binaries find that dynamical heating and cluster expansion strongly suppress late-time mergers, making detectable rates implausible in that specific configuration, and are performed assuming a monochromatic PBH mass distribution \citep{10.1093/mnras/staa1644}. Conversely, cosmological collisional  simulations that include continuous accretion, many-body interactions, and hierarchical structure formation adopt an extended PBH mass function and find that late binaries can form and merge, with coalescences driven stochastically by encounters and eccentricity excitation rather than secular hardening \citep{Delos_2024} \footnote{We note that \cite{Delos_2024} also find that the cumulative GW background from PBH binaries may exceed current LVK limits; however, their estimate depends on the assumed PBH mass function, which differs from the extended distribution adopted in this work, and a detailed assessment of the resulting gravitational-wave background is beyond the scope of this paper.}. These results demonstrate that late-channel merger rates are highly environment-dependent and analytically unreliable, rather than universally negligible. In this context, our predicted merger rate should be interpreted as an illustrative estimate within a standard late-capture framework, not as a precision constraint. Importantly, the primary aim of this work is to assess whether the hypothesized sub-solar-mass merger rate inferred from one LVK detection is broadly compatible with PBH scenarios under commonly adopted astrophysical assumptions.
\section{Detected event rate}
\label{sec:detected_rate}
The detector response as a function of frequency is encoded in the one--sided strain noise power spectral density $S_h(f)$, derived from the amplitude spectral density ${\rm ASD}(f)$ via $S_h(f) = [{\rm ASD}(f)]^2$.
We used O3b H1 LVK amplitude spectral density, publicly available at \url{https://dcc.ligo.org/LIGO-G2100672/public}.
The detector horizon distance $r_{\rm det}(M_1,M_2)$, which is the luminosity distance of the farthest detectable source, can be expressed in the following form \citep{Carr2021-lr}:
\begin{align}
\begin{split}
    &r_{\rm det}(M_1,M_2)
    = \frac{\sqrt{5}\;c}{24\,\pi^{2/3}\,2.26}
      \left(\frac{G \mathcal{M}_c}{c^3}\right)^{5/6}
      \, \times \\
    &\times\left[
        \int_{f_{\min}}^{f_{\rm ISCO}(M_1,M_2)} 
          \frac{f^{-7/3}}{S_h(f)}\,{\rm d}f
        + \int_{f_{\rm ISCO}(M_1,M_2)}^{f_{\max}(M_1,M_2)}
          \frac{f^{-2/3}}{S_h(f)}\,{\rm d}f
      \right]^{1/2}
    \label{eq:detector}
\end{split}
\end{align}
where $\mathcal{M}_c = \frac{(M_1 M_2)^{3/5}}{(M_1+M_2)^{1/5}} \,$ is the chirp mass and the 2.26 factor is for an Euclidean Universe. In principle, one should account for redshift eﬀects on the waveform by replacing $\mathcal{M}_c$ by $\mathcal{M}_c(1 + z)$. However, we argue that such a detector probes the local Universe, and the redshift effects for low mass BH binaries are subdominant compared to the intrinsic astrophysical uncertainties of the model and the main inaccuracies outlined in \cite{Carr2021-lr}. For such low-mass binaries the detectable volume lies at $z \lesssim 0.05$, so cosmological corrections to the waveform and distance are negligible at the few-percent level.

The square brackets describe the contributions over the frequency bandwidth of the detector: the first contribution accounts for the post-Newtonian waveform of the inspiral phase  $f^{-7/3}$ between the minimum frequency detectable from the interferometers ($f_{min}=10 \;\mathrm{Hz}$ for LVK O3b) and $f_{ISCO}= \frac{4400}{M_{tot}} \;\mathrm{Hz}$ with $M_{tot}$ in solar masses; the second piece includes the merging phase with a frequency dependence of $f^{-2/3}$ between $f_{ISCO}$ and $f_{max}=\mathrm{min}(f_{merge},2000 \;\mathrm{Hz})$, with the merge frequency $f_{merge}$ given by \citep{Ajith_2008}
$\pi M_{tot} f_{\rm merge} = a_0 \eta^{2} + b_0 \eta + c_0$,
where 
$\eta = \dfrac{M_1 M_2}{(M_1 + M_2)^2}$
is the symmetric mass ratio of the binary and the coefficients for $f_{\rm merge}$ are $
    a_0 = 6.6389 \times 10^{-1},
    b_0 = -1.0321 \times 10^{-1},
    c_0 = 1.0979 \times 10^{-1}$.

The comoving volume effectively surveyed for that binary is 
\begin{equation}
    V_{\rm det}(M_1,M_2)
    = \frac{4\pi}{3}\,r_{\rm det}^3(M_1,M_2)\,
\end{equation}
so the detected merger rate for binaries with component masses $(M_1,M_2)$ is then
\begin{equation}
    R_{\rm det}(M_1,M_2)
    = \tau_L(M_1,M_2)\,V_{\rm det}(M_1,M_2).
\end{equation}
To keep the analysis light and consistent with S251112cm, we imposed a selection cut on the mass ratio, defined as $q(M_1,M_2) \equiv \frac{\min(M_1,M_2)}{\max(M_1,M_2)}$, and the chirp mass, such that $q_{\min}=0.05 \leq q(M_1,M_2) \leq q_{\max}=1$
and $\mathcal{M}_{c,\min}=0.1 \leq \mathcal{M}_c/M_\odot \leq \mathcal{M}_{c,\max}=0.87.$

The total detected rate is then:
\begin{equation}
    R_{\rm tot}
    = \sum_{i=1}^{N_M}
      \sum_{j=1}^{i}
      \Theta_{\rm sel}(M_i,M_j)\,
      R_{\rm det}(M_i,M_j) \,,
\label{eq:R_tot}
\end{equation}
where $\Theta_{\rm sel}(M_1,M_2)$ is a selection function equal to 1 when all the cuts are satisfied at the same time, and 0 otherwise.
Equivalently, in the continuous limit one may write
\begin{equation}
    R_{\rm tot}
    = \frac{1}{2}\iint 
      {\rm d}\ln M_1\,{\rm d}\ln M_2\,
      \Theta_{\rm sel}(M_1,M_2)\,
      R_{\rm det}(M_1,M_2)\,,
\end{equation}
with $R_{\rm det}$ defined as above and the number densities encoded via the PBH mass function.
\begin{figure*}
    \centering
    \includegraphics[width=\linewidth]{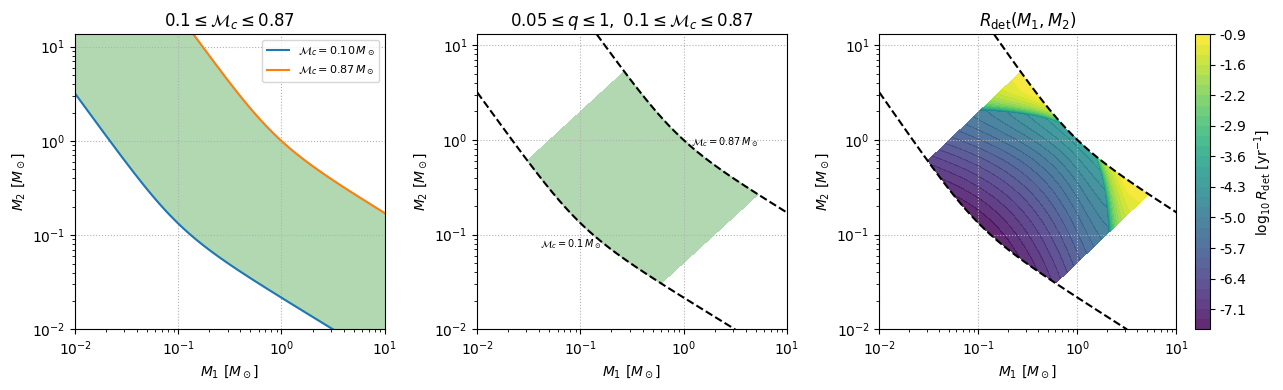}
    \caption{Left panel: region in the \((M_1,M_2)\) plane satisfying \(0.1 \leq \mathcal{M}_c(M_1,M_2) \leq 0.87\). Central panel: subset of this region further restricted to mass ratios \(0.05 \leq q(M_1,M_2) \leq 1\). Right panel: expected detectable merger rate across the resulting allowed parameter space. Note that in this figure we did not enforce \(M_1 > M_2\); however, when computing the total rate \(R_{\rm tot}\) we summed only the contributions with \(M_1 > M_2\), i.e. over the half--plane corresponding to ordered pairs.
}
    \label{fig:m1_vs_m2}
\end{figure*}

\section{Discussion}
\label{sec:discussion}

Using the methodology detailed in Sections \ref{sec:merger_rate} and \ref{sec:detected_rate}, we calculated the total detectable event rate ($R_{\rm tot}$) of PBH binaries by assuming the \cite{Bodeker2021-hz} extended PBH mass function modified by lepton-flavour asymmetries (the solid blue line in Fig.~\ref{fig:massfunction}), which assumes a total dark matter fraction of $f_{\rm PBH} = 0.339$ in the considered mass range and is consistent with existing accretion, microlensing and GW constraints (right panel, Fig.~\ref{fig:massfunction}).

In Fig.~\ref{fig:m1_vs_m2} we show the detectable merger rate $R_{\rm det}(M_1, M_2)$ as a function of the component masses within the allowed region defined by $\mathcal{M}_c \in [0.1, 0.87] M_{\odot}$ and a mass ratio $q \in [0.05, 1]$. The plot  shows that the detectable rate is maximized for binaries with relatively high $q$ in the range of approximately $0.1 - 5.0 M_{\odot}$. 

By integrating the detectable rate over this allowed parameter space (Eq. \ref{eq:R_tot}), we obtain the total predicted detected event rate for an LVK-like detector sensitivity (O3b H1) of $R_{\rm tot} = 0.8 \text{ yr}^{-1}$.
This detectable rate corresponds to an intrinsic local merger rate density (in the parameter space analyzed) of $\mathcal{R}_{0.1-0.87} \approx 3.45 \text{ Gpc}^{-3} \text{yr}^{-1}$.
Our predicted detectable rate of $R_{\rm tot} = 0.8\ \text{yr}^{-1}$ lies well within the 95\% C.L.\ interval derived from one single detection from LVK. For one detected event, the 95\% single-sided Poisson limits on the true mean rate are $\lambda_{\rm low} = 0.0513$ and $\lambda_{\rm up} = 4.74$ \citep{1986ApJ...303..336G}. Over the 4.35 effective years during which the LVK network was operational \textemdash approximated here with O3b-like sensitivity across all runs, as O1--O2 were less sensitive and O4 more sensitive, making O3b a reasonable midpoint approximation \textemdash the corresponding rate is $R_{\rm SingleDetection} = 0.23\ \rm yr^{-1}$. This yields an inferred sub-solar merger rate of
$0.012\ {\rm yr^{-1}} < R_{\rm obs} < 1.09\ {\rm yr^{-1}} \; (95\%~{\rm C.L.})$, equivalent to one detection every $0.9$ to $83$ years.

This suggests that a PBH population formed during the QCD epoch is not only a viable but a compelling interpretation for one LVK sub-solar detection during the first four runs.

However, the late-time capture rate (Eq. \ref{eq:tau_L}) scales strongly with the local astrophysical parameters, primarily as $\propto \delta_{\rm loc}^2 v^{-11/7}$. We adopted standard average values for a Milky Way-like halo ($\delta_{\rm loc} \sim 3\cdot10^5$, $v \sim \sqrt{2}\times v_{vir},\; v_{vir} \sim250 \text{ km s}^{-1}$), but the true cosmic-averaged values are subject to significant astrophysical uncertainties and may therefore impact the value of late-time capture rates.
Furthermore, our rate calculation scales with the assumed PBH abundance squared, $f_{\rm PBH}^2$.

We can, however, reverse the argument: instead of assuming $f_{\rm PBH}$ to predict a rate, we can use the \emph{observed} rate to constrain the PBH abundance. The fact that our $R_{\rm tot}$ (based on $f_{\rm PBH} = 0.339$) aligns so well with $R_{\rm obs}$ suggests that an $f_{\rm PBH}$ of this order is plausible. 
To convert one single sub-solar detection into a constraint on the PBH abundance, we use the fact that, for late--Universe gravitational capture, the detectable merger 
rate scales as $R_{\rm det} \propto f_{\rm PBH}^{2}$.
Relative to the fiducial model used 
in this work, which predicts $R_{\rm ref} = 0.8~{\rm yr^{-1}}$ for $f_{\rm PBH,ref}=0.339$,
the abundance required to match any observed rate $R_{\rm obs}$ is $f_{\rm PBH,low}=f_{\rm PBH,ref}\,\sqrt{\frac{R_{\rm obs}}{R_{\rm ref}}}$.
Using the single--event 95\% Poisson lower limit for the true mean number of
events, $\lambda_{\rm low}=0.0513$, and the effective observing time of the LVK network used 
in this analysis ($T=4.35$ yr), the corresponding lower bound on the merger rate is $R_{\rm obs,low}=0.012 \;\rm yr^{-1}$. 
Substituting this value into the scaling relation yields
$f_{\rm PBH,low}
=0.339\,\sqrt{\frac{1.18\times10^{-2}}{0.8}}
\simeq 0.04.$
Therefore, one sub-solar GW event, if ever detected and confirmed, places a conservative, statistically consistent lower limit on the primordial-black-hole abundance of
${f_{\rm PBH}>0.04}$. We note that this result is based on our mass function and is sensitive to halo modeling assumptions.

%We must consider however that sub-solar mass compact objects forming via standard stellar evolution is possible, even if disfavored. The primary astrophysical alternative would involve at least one neutron star \cite{???}. However, the LVK analysis itself reported a low probability ($<8\%$) of a neutron star being present. This classification is strongly corroborated by the extensive, multi-wavelength electromagnetic follow-up campaigns \citep{2025GCN.42707....1M,2025GCN.42666....1G,2025GCN.42675....1F}, which found no credible kilonova or other EM counterpart. Given the extreme challenges of stellar-mass formation channels and the conspicuous lack of EM emission, the PBH hypothesis stands as the simplest explanation for S251112cm.

The low but non-zero value of $R_{\rm tot}$ is entirely consistent with the observed rarity of GW events in the sub-solar search window, suggesting that S251112cm might constitute the inaugural astrophysical detection of a sub-solar mass PBH binary, lending crucial support to the existence of a PBH population originating from the QCD epoch as a significant, if not total, contributor to dark matter.
While the PBH formation channel provides a robust interpretation for the sub-solar chirp mass (in case the final mass parameter estimation confirms these numbers), alternative astrophysical mechanisms must be considered. Specifically, disk fragmentation or core fissioning during stripped-envelope supernovae (SESNe) has been theoretically suggested as a pathway that could lead to the formation of sub-solar mass neutron stars \citep{https://doi.org/10.48550/arxiv.2508.17183_SESNe}, which could subsequently merge \citep{PhysRevD.107.103012}. Although the probability of the event containing a neutron star is low ($<8\%$), the presence of a neutron star would inherently connect the event to heavy element nucleosynthesis pathways (such as the r-process). The rapid electromagnetic follow-up conducted by numerous collaborations \citep[see e.g.,][]{42677,42658}
implicitly investigates these possibilities, as kilonovae are expected counterparts to mergers involving a neutron star. 
As of the time of writing, extensive rapid electromagnetic follow-up across the optical (GOTO, BlackGEM/MeerLICHT, ATLAS) and high-energy (Fermi-GBM, Swift XRT) regimes placed strong upper limits on any associated emission in the ${1681 \text{ deg}^2}$ credible region \citep{S251112cm}. In the absence of any credible transient that can be unambiguously linked to a kilonova counterpart \textemdash with all vetted candidates either identified as supernovae or inconsistent with the distance constraints \citep{2025GCN.42666....1G, 2025GCN.42675....1F} \textemdash S251112cm remains most plausibly classified as a compact-object merger unlikely to contain a neutron star. 

The absence of an electromagnetic counterpart does not uniquely favor a PBH origin, as other compact-object mergers are intrinsically electromagnetically dark. 
More exotic compact objects could in principle produce electromagnetically dark mergers in this mass range; however, unlike primordial black holes, such scenarios currently lack well-established formation mechanisms that naturally predict both the relevant mass scale and the implied merger rate.

As an additional consistency check, we applied the same late-Universe capture and detectability
procedure described above to the stellar-mass domain probed by LVK, restricting the component
masses to $3 \le M/M_\odot \le 200$. Integrating the detected rate over this interval yields a total
predicted detection rate of
$R^{3-200}_{\rm tot} \simeq 120\ {\rm yr^{-1}} $.
Current scenarios produce markedly different LVK detection rates; for instance, for O3 the
overall predicted rates span $\sim 22.2$ to $148.3\ {\rm yr^{-1}}$ across representative models
\citep{drq9-dpy4_LVKprediction}. Our value lies within this range, indicating that this result remains compatible with a non-negligible fraction of the BBH mergers observed
by LVK in the $3$-$200\,M_\odot$ range having a primordial origin.

\section{Summary}
\label{sec:summary}

The LVK collaboration has reported S251112cm, a compact binary merger candidate with a chirp mass $\mathcal{M}_c$ in the range [0.1, 0.87] $M_{\odot}$ with at least one of the component masses in the sub-solar mass gap, a region where black holes of stellar origin are not expected to form.

In this paper, based on the S251112cm trigger, we tested the hypothesis that a population of Primordial Black Holes formed during the QCD epoch could be seen by LVK, using a physically motivated, extended mass function modified by lepton-flavour asymmetries (Fig.~\ref{fig:massfunction}). We calculated the expected detectable merger rate ($R_{\rm tot}$) for the LVK O3b run, assuming binaries form via the dominant late-time gravitational capture channel in dark matter halos.

To summarize, our results suggest that:
a) Our model, using an $f_{\rm PBH}=0.339$ (in this mass range) and standard halo parameters, predicts a detectable sub-solar mass event rate of $R_{\rm tot} = 0.8 \text{ yr}^{-1}$.

b) This \emph{ab initio} prediction is in good agreement with the rate inferred from  one single detection during the first four runs of LVK, which is $R_{\rm obs}\simeq 0.23^{+0.86}_{-0.218} \text{ yr}^{-1}$ (95\% C.L.).
 
%b) The parameters of S251112cm fall precisely in the mass range (see Fig.~\ref{fig:m1_vs_m2}, right panel) that is maximally sensitive to the sub-solar peak of our modeled QCD-epoch PBH mass function.

We conclude that S251112cm, if validated, is a compelling candidate for the first detection of a merging sub-solar mass PBH binary. If confirmed as astrophysical, the inferred sub-solar mass scale and the absence of evidence for neutron-star involvement would be difficult to reconcile with standard stellar formation channels; primordial black holes provide a physically motivated scenario consistent with these properties and with the inferred merger rate.

Apart from the inferred component-mass range, which reflects the LVK parameter-estimation pipeline, the specific observational details of S251112cm are not directly relevant to the merger-rate calculation presented here. The rate estimate depends only on the existence of a confident detection of a sub-solar-mass compact binary and on the corresponding spacetime volume probed by the detectors, rather than on event-specific quantities such as sky localization, distance, or signal morphology. Consequently, the same reasoning would apply to any confirmed detection of a sub-solar-mass binary, and our analysis should be interpreted as assessing the implications of such a detection in general, rather than being tied to the particular properties of S251112cm.

This single event, if its astrophysical origin is confirmed by future analysis, has profound implications. It transitions the sub-solar mass region from a theoretical curiosity to an observational frontier. It places a new lower limit on the PBH contribution to dark matter, $f_{\rm PBH} > 0.04$, assuming our mass function and the astrophysical parameters selected. Extending this, if a fraction of the heavier, stellar-mass BBHs detected by LVK also have a primordial origin (as predicted by our broad mass function), it is plausible that PBHs constitute a significant fraction of all dark matter, $f_{\rm PBH} \sim 0.339$, in the selected mass range. The forthcoming LVK O5 observing run will provide the crucial statistics needed to test this hypothesis and potentially unveil the nature of dark matter.
\begin{acknowledgements}
N.C. dedicates this work to his beloved father, \textit{Vincenzo Cappelluti}, who departed this planet during the drafting of this manuscript. His memory fills the void he left, as completely as the energy that permeates the Universe.

The authors kindly acknowledge the LIGO-Virgo-KAGRA collaboration for the prompt publication of the event discovery notice. The authors thank Alberto Salvarese for bringing this gravitational-wave event to their attention, which prompted part of the analysis presented in this work. The authors acknowledge the University of Miami for partial support.
\\ \textit{Software:} The analysis has been performed by using the \href{https://www.python.org/}{Python} \citep{10.5555/2011965Python} programming language  and \href{https://jupyter.org/}{Jupyter Notebook} \citep{Kluyver:2016aajupyter} interactive computational environment. Specifically, the following packages have been employed: \href{https://numpy.org/}{Numpy} \citep{harris2020arrayNumpy}, \href{https://matplotlib.org/stable/}{Matplotlib} \citep{Hunter:2007matplotlib}, \href{https://scipy.org/}{Scipy} \citep{2020SciPy-NMeth}, \href{https://pandas.pydata.org/}{Pandas} \citep{the_pandas_development_team_2025_17229934}.
\end{acknowledgements}

\bibliography{sample701}{}
\bibliographystyle{aasjournalv7}
\end{document}